\documentclass[preprint,review,12pt]{elsarticle}

\usepackage{graphicx}
\usepackage{amsmath}
\usepackage{amssymb}

\journal{Physics Letters B}

\begin{document}

\begin{frontmatter}

\title{Fragmentation of spin-dipole strength in $^{90}$Zr and $^{208}$Pb}

\author[GSI,Zagreb]{T. Marketin}

\author[EMMI]{E. Litvinova}

\author[Zagreb]{D. Vretenar}

\author[Munich]{P. Ring}

\address[GSI]{GSI Helmholtzzentrum f\"{u}r Schwerionenforschung, Planckstra\ss e 1, D-64291 Darmstadt, Germany}
\address[Zagreb]{Physics Department, Faculty of Science, University of Zagreb, 10000 Zagreb, Croatia}
\address[EMMI]{ExtreMe Matter Institute EMMI and Research Division, GSI Helmholtzzentrum f\"{u}r Schwerionenforschung, Planckstra\ss e 1, D-64291 Darmstadt, Germany}
\address[Munich]{Physik-Department der Technischen Universit\"{a}t M\"{u}nchen, D-85748 Garching, Germany}

\begin{abstract}
An extension of time-dependent covariant density functional theory that includes
particle-vibration coupling is applied to the charge-exchange channel. 
Spin-dipole excitation spectra are calculated an compared to available data for $^{90}$Zr and
$^{208}$Pb. A significant fragmentation is found for all three 
angular-momentum components of the spin-dipole strength as a result of 
particle-vibration coupling, as well as a shift of a portion of the strength to higher energy. 
A high-energy tail is formed in the strength distribution that  
linearly decreases with energy. Using a model-independent sum rule,
the corresponding neutron skin thickness is estimated and shown to be consistent 
with values obtained at the mean-field level.
\end{abstract}


\end{frontmatter}

Spin-isospin excitations present a very active research 
topic both in nuclear structure and nuclear astrophysics. 
In particular, a detailed knowledge of the Gamow-Teller resonance, 
a collective oscillation of excess neutrons that coherently change the direction 
of their spin and isospin without changing their orbital motion, is essential
for understanding weak nuclear reactions involved in the process of
nucleosynthesis, i.e. $\beta$-decay, electron and neutrino capture.
Moreover, it has been shown that spin-dipole charge-exchange excitations, 
made up of three components with angular momentum and parity $J^{\pi} =
0^{-}$, $1^{-}$ and $2^{-}$, can significantly contribute to
the total reaction rates and even compete with the contribution
of Gamow-Teller transitions~\cite{Kolbe2001a,Suzuki2003,Paar2008}.

The spin-dipole strength can also provide information on basic 
properties of finite nuclei. The thickness of the neutron skin has been
shown to constrain the neutron equation of state~\cite{Typel2001},
and is also correlated with the nuclear symmetry
energy~\cite{Yoshida2006}. While accurate data on the charge
distribution in nuclei have been obtained by elastic electron
scattering, our knowledge of neutron distribution comes primarily from hadron
scattering, and the results are markedly model-dependent. Indirect
methods for determining the neutron skin thickness have been
proposed, based on energy differences between the Gamow-Teller
and isobaric analogue resonances~\cite{Vretenar2003}, and using the
model-independent sum rule for the spin-dipole
resonance~\cite{Krasznahorkay1999}. Two
recent $(p,n)$ and $(n,p)$ measurements of the spin-dipole response
of $^{90}$Zr and $^{208}$Pb~\cite{Yako2006,Wakasa2010} have also 
prompted new theoretical studies, in particular investigations based on the random phase
approximation~\cite{Sagawa2007, Fracasso2007, Liang2008, Bai2010}.

Both measurements \cite{Yako2006,Wakasa2010} show a high-energy tail in the 
spin-dipole strength distribution that cannot be described by the simple 
one-particle -- one-hole
(1p1h) random phase approximation (RPA). Previous attempts to extend 
this framework using the 2p2h RPA were based on a  
non self-consistent approach that employs a phenomenological Woods-Saxon
potential to obtain the ground-state wave functions~\cite{Drozdz1987}. 
In this Letter we introduce a charge-exchange version of the particle-vibration coupling model
based on time-dependent covariant density functional theory, and apply it to an 
analysis of spin-dipole strength distributions in $^{90}$Zr and $^{208}$Pb.

The basic quantity that describes small-amplitude motion of an
even-even nucleus in an external field with frequency $\omega$
is its response function $R(\omega)$ \cite{Ring1980}. It is obtained as 
a solution of the Bethe-Salpeter equation:
\begin{equation}
R(\omega) = {\tilde R}^{0}(\omega) + {\tilde R}^{0}(\omega) W(\omega)R(\omega),
\end{equation}
where ${\tilde R}^{0}(\omega)$ is the propagator of two
uncorrelated quasiparticles in the static mean field, and the second
term includes the in-medium nucleon-nucleon interaction
$W(\omega)$. The two-body interaction $W(\omega)$ contains
static terms and a frequency-dependent term:
\begin{equation} \label{eq:inter}
W(\omega) = V_{\rho} + V_{\pi} + V^{LM}_{\delta\pi} + \Phi(\omega) - \Phi(0).
\end{equation}
$V_{\rho}$ and $V_{\pi}$ represent the finite-range $\rho$-meson and $\pi$-meson 
exchange interactions, respectively. They are derived from the covariant energy 
density functional and read:
\begin{eqnarray}
V_{\rho}(1,2) & = & g_{\rho}^2{\vec\tau}_1{\vec\tau}_2(\beta\gamma^{\mu})_1(\beta\gamma_{\mu})_2 D_{\rho}({\bf r}_1,{\bf r}_2), \nonumber\\
V_{\pi}(1,2) & = & \Bigl(\frac{f_{\pi}}{m_{\pi}}\Bigr)^{2}{\vec\tau}_1{\vec\tau}_2({\bf\Sigma}_1{\bf\nabla}_1)
({\bf\Sigma}_2{\bf\nabla}_2)D_{\pi}({\bf r}_1,{\bf r}_2),
\end{eqnarray}
where $g_{\rho}$ and $f_{\pi}$ are the coupling strengths,
$D_{\rho}$ and $D_{\pi}$ are the meson propagators and ${\bf\Sigma}$
is the generalized Pauli matrix~\cite{Paar2004}.  The derivative
type of the pion-nucleon coupling necessitates the inclusion of the
zero-range Landau-Migdal term $V^{LM}_{\delta\rho}$, which accounts
for the contact part of the nucleon-nucleon interaction 
\cite{Paar2004}. $\Phi(\omega)$ describes the coupling of the
quasiparticles to vibrations (phonons) generated by coherent
nucleonic motion. In the quasiparticle time blocking approximation
(QTBA) \cite{Tselyaev2007} it can be written in the following
operator form:
\begin{equation} \label{eq:phi}
\Phi(\omega) = \sum\limits_{m,\eta}g_m^{(\eta)\dagger}{\tilde R}^{0(\eta)}(\omega - \eta\omega_m)g_m^{(\eta)},
\end{equation}
where the index $m$ enumerates vibrational modes with frequencies
$\omega_m$ and coupling amplitude matrices $g_m^{(\eta)}$, and the
index $\eta = \pm 1$ denotes forward and backward components. In 
Eq. (\ref{eq:inter}) the term $\Phi(0)$ is subtracted to remove the 
effect of double counting the phonon coupling, because the
parameters of the density functional have been adjusted to 
ground-state data and, therefore, already include
essential static phonon contributions. The energy-dependent
effective interaction of Eq. (\ref{eq:phi}) leads to 
fragmentation of nuclear spectra and determines the width
of  giant resonances \cite{Litvinova2007}.

The strength function $S(\omega)$:
\begin{equation} \label{eq:strf}%
S(E,\Delta )=-\frac{1}{\pi} Im \langle P^{\dagger}R(E+i\Delta)P\rangle,
\end{equation}
yields the spectral distribution of the nuclear response in a
given external field $P$. The field operators for 
charge-exchange spin-dipole transitions read: 
\begin{equation} \label{eq:operator}
{P}_{\pm}^{\lambda} = \sum_{i} r(i) \left[ \boldsymbol{\sigma}(i)
\otimes Y_{1}(i) \right]_{\lambda} t_{\pm}(i),
\end{equation}
where $t_{\pm}$ denotes the isospin raising and lowering operators.

\begin{figure}[htb]
\centerline{ 
  \includegraphics[width=100mm,angle=0]{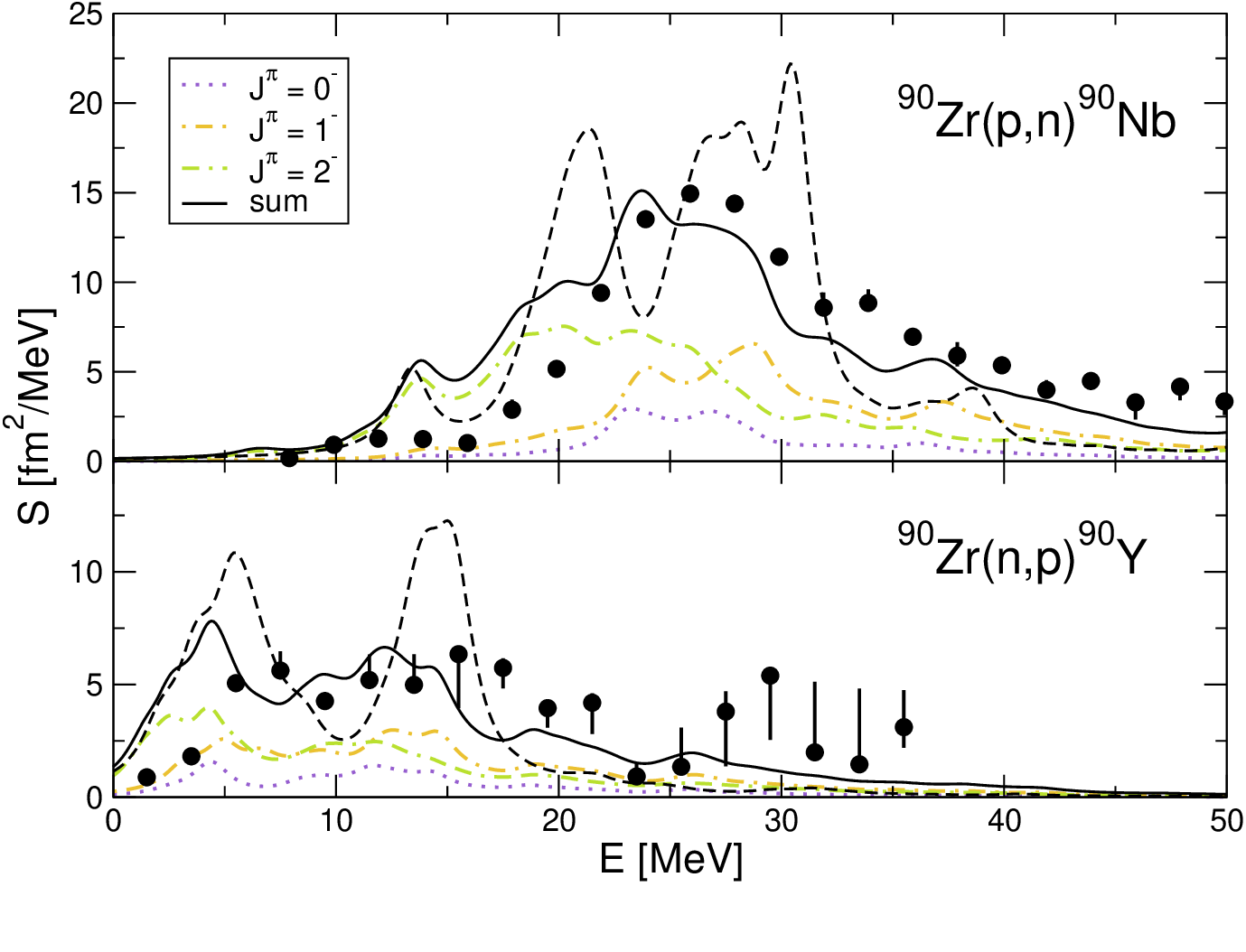}%
} 
\caption{\label{fig:Zr} Spin-dipole strength
distributions for the t$_{-}$ (upper panel) and the t$_{+}$ (lower
panel) channels in $^{90}$Zr. On the horizontal axis the excitation
energy is plotted with respect to the ground state of $^{90}$Zr. The solid black
curve represents the sum of the strength distributions of the
$0^{-}$, $1^{-}$ and $2^{-}$ components, calculated in the RTBA
that includes particle-vibration coupling. The dashed
black curves denotes the corresponding RRPA strength function. In both
cases the imaginary part of the energy is set to $\Delta = 1$ MeV.
The experimental results denoted by full circles are from Ref.
\cite{Yako2006}.}
\end{figure}

The phonon coupling terms augment the RPA spectrum with additional
p-h$\otimes$phonon components that generally lead to significant
fragmentation of giant resonances~\cite{Litvinova2007}. In
Fig.~\ref{fig:Zr} we compare the spin-dipole strength distribution in $^{90}$Zr
calculated using the relativistic random phase approximation (RRPA), 
and the relativistic time-blocking approximation (RTBA). In both models 
the NL3 \cite{LKR.97} relativistic mean-field effective interaction has consistently 
been used for the calculation of the mean-field ground-state, the RPA phonons, 
and the spin-dipole charge-exchange excitations. 

The prominent RRPA peaks (dashed curves) disappear when 
particle-vibration coupling is included in the RTBA (solid curves).
In the $t_{-}$ channel, for instance, only a broad
resonance remains with the peak at 23.5 MeV, in very good 
agreement with available data~\cite{Yako2006}.
The three angular-momentum components do not, however, 
follow the energy hierarchy $E(2^{-}) < E(1^{-}) < E(0^{-})$ 
predicted by recent Skyrme-RPA calculations~\cite{Sagawa2007, Fracasso2007}. 
While the RTBA predicts the $2^{-}$ component to be the lowest one with the centroid
energy $m_{1}^{2^{-}}/m_{0}^{2^{-}} = 25.4$ MeV, $0^{-}$ is
found to be lower than the $1^{-}$ component, with centroid
energies at 29.4 MeV and 32.6 MeV, respectively. This is probably 
due to the fact that the exchange terms are neglected in the mean-field
calculation. Namely, as shown in a recent study  \cite{Liang2008}, a fully 
consistent relativistic Hartree-Fock (RHF) + RPA 
calculation yields  $E(2^{-}) < E(1^{-}) < E(0^{-})$ for the excitation energies 
of spin-dipole components.

The inclusion of particle-vibration coupling leads to a shift of the strength to
higher excitation energies. A high-energy tail is formed in the region above 
30 MeV where the strength decreases almost
linearly with increasing energy, in close agreement with experimental results. 
In contrast, the RRPA strength decreases more rapidly
above 30 MeV, and becomes 5 to 10 times smaller than the experimental 
strength above 40 MeV.

In the $t_{+}$ channel the two dominant peaks predicted by the RRPA
merge into a single broad structure that extends up to approximately
15 MeV excitation energy. The tail at higher energies decreases 
approximately linearly with increasing energy. One might notice a very
good agreement with data, except in the low-energy region below 5 MeV, 
where both the RRPA and the RTBA predict spin-dipole strength, 
originating predominantly from the $2^{-}$ component, that is 
considerably larger than the measured distribution.

\begin{figure}
\centerline{
  \includegraphics[width=100mm,angle=0]{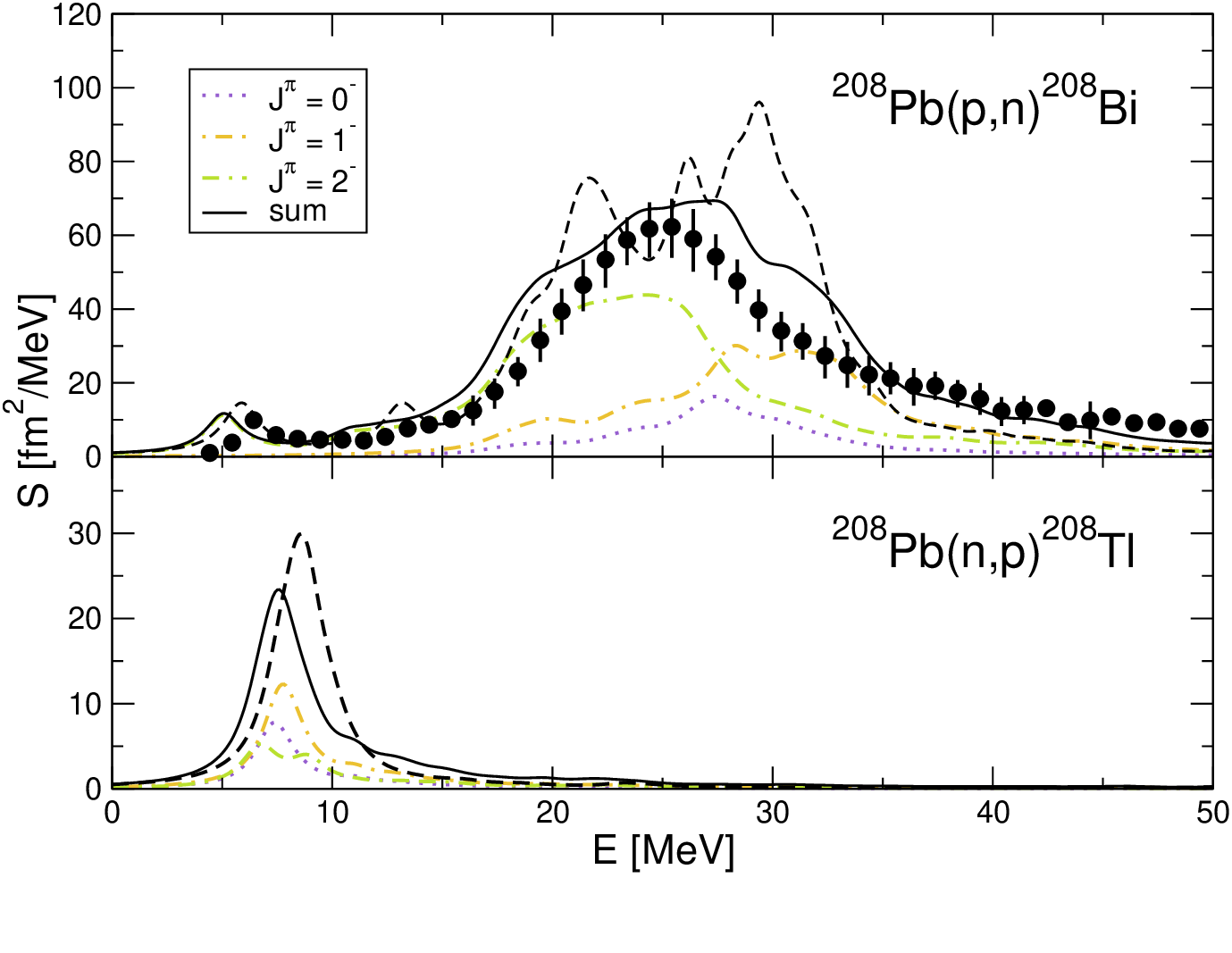}%
} \caption{\label{fig:Pb} Same as described in the caption to Fig. \ref{fig:Zr}
but for $^{208}$Pb. The data are from Ref.
\cite{Wakasa2010}.}
\end{figure}

The results for the spin-dipole strength in $^{208}$Pb are shown in Fig.~\ref{fig:Pb}.
The data exhibit a broad asymmetric resonance centered at 25
MeV, and an additional small peak at approximately 6 MeV. 
The RTBA results reproduce these structures, even though 
the calculated width of the main resonance is
slightly larger than the empirical value. As in the previous case, 
a portion of the strength is shifted to higher energies by the 
inclusion of particle-vibration coupling, in very good  agreement 
with data above 35 MeV. Obviously in this region the RRPA 
strength distribution decreases much faster with energy compared 
to RTBA. The ordering of the angular-momentum components of the
strength is the same as in the case of $^{90}$Zr: $E(2^{-}) <
E(0^{-}) < E(1^{-})$. The lower
panel of Fig.~\ref{fig:Pb} displays the distributions of the SD$_{+}$ strength. 
In this case the strength is concentrated in a single peak
centered at 7.5 MeV. Relatively little fragmentation is obtained in
comparison with the RRPA results, even though some strength is shifted 
to higher excitation energy in the RTBA. These findings are supported by the
available data from Ref.~\cite{Long1998}, where a single peak has
been observed at approximately 7.5 MeV excitation energy.

\begin{figure}
\centerline{
  \includegraphics[width=100mm,angle=0]{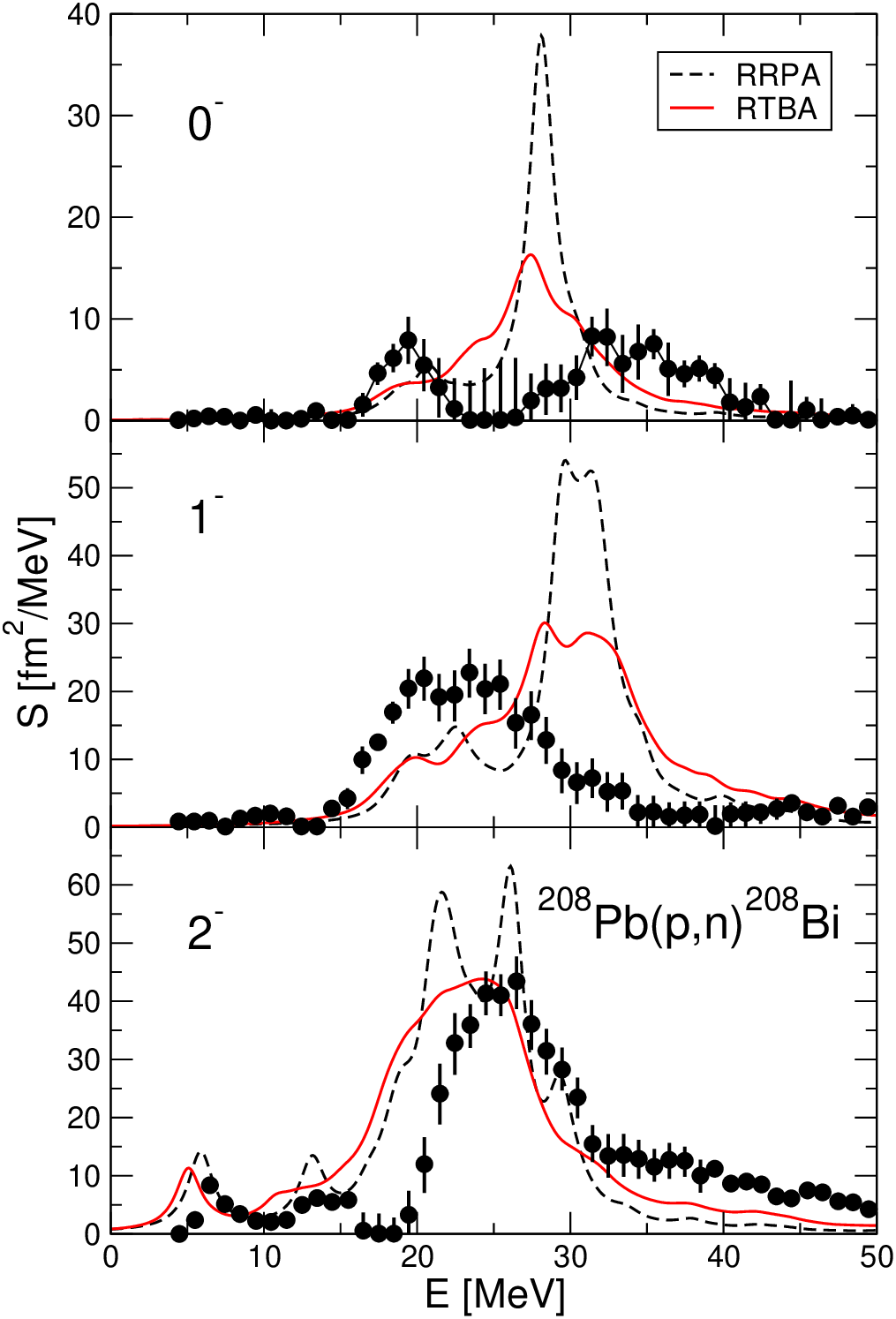}%
} \caption{\label{fig:PbMultipoles} Spin-dipole
strength distributions in $^{208}$Pb for the $J^{\pi} = 0^{-}, 1^{-}$
and $2^{-}$ components. The black dashed curves are the RRPA results, 
and the solid red curves denote results
obtained with the particle-vibration coupling model. Data are
from Ref. \cite{Wakasa2010}.}
\end{figure}

In the case of $^{208}$Pb data are also available for
each component of the spin-dipole strength~\cite{Wakasa2010}. In
Fig.~\ref{fig:PbMultipoles} we display a comparison between the 
RRPA, RTBA and the experimental results for $J^{\pi} = 0^{-}$, $1^{-}$ and
$2^{-}$. The RRPA predicts that the strength of the $0^{-}$ component is
concentrated in a single peak at the excitation energy of $\approx 28$ MeV with
respect to the ground state of $^{208}$Pb. Particle-vibration
coupling induces fragmentation and spreading of this strength, but
the basic structure of the distribution is not altered. Obviously 
this does not completely agree with the experimental results. 
For the $1^{-}$ components the main peak is centered around 30 MeV,
whereas the experiment places it around 23 MeV. The opposite situation
occurs for the $2^{-}$ component, for which the calculated distribution is 
in qualitative agreement with experiment, even though the centroid of the 
main peak is calculated few MeV below the measured resonance. 
In Ref.~\cite{Bai2010} the results have been brought in agreement with 
experiment by the inclusion of tensor correlations in the
Skyrme energy density functional. These correlations exhibit a
multipole-dependent effect on spin-dipole excitations. It is
interesting to note that in this case experimental results indicate 
that the $1^{-}$ component of the spin-dipole resonance is actually 
below the $2^{-}$ component.

Using the calculated strength distributions, we have also determined the sum-rule values 
for the spin-dipole response of $^{90}$Zr and $^{208}$Pb. 
The model independent sum rule relates the spin-dipole strength to the neutron and proton 
ground-state radii~\cite{Gaarde1981}:
\begin{equation} \label{eq:sumrule}
S_{-}^{\lambda} - S_{+}^{\lambda} = \frac{2\lambda + 1}{4\pi} \left( N \left\langle r^{2} \right\rangle_{n} - Z \left\langle r^{2} \right\rangle_{p} \right),
\end{equation}
where $S_{\pm}^{\lambda}$ denotes the total SD strength in the $t_{\pm}$ channel  
for angular momentum $\lambda$. 
$S_{\pm}$ will denote the sum of spin-dipole strengths of the three components. 
The neutron skin thickness $\delta_{np}$ is defined as the difference of neutron and proton $rms$ radii: 
\begin{equation} \label{eq:delta}
\delta_{np} = \sqrt{\left\langle r^{2} \right\rangle_{n}} - \sqrt{\left\langle r^{2} \right\rangle_{p}}.
\end{equation}
\begin{table}
\caption{\label{tab:nskin} Spin-dipole sum-rule values, proton and
neutron $rms$ radii, and thickness of the neutron skin of $^{90}$Zr and
$^{208}$Pb. The proton radii correspond to the self-consistent
ground-state distributions, whereas the neutron radii are calculated using
the sum rule of Eq. (\ref{eq:sumrule}). Sum rules are given in
units of fm$^{2}$, and radii in fm. Error estimates in the experimental value of the sum 
rule arise from statistical, systematic and multipole decomposition uncertainties, respectively.}
\begin{center}
\begin{tabular}{c|cc}
\hline
 & $^{90}$Zr & $^{208}$Pb \\
\hline
$S_{-} - S_{+}$ (g.s.) & 160.925 & 1222.044 \\
$S_{-} - S_{+}$ (calc.)& 160.963 & 1213.562 \\
$S_{-} - S_{+}$ (exp.) & $148 \pm 6 \pm 7 \pm 7$~\cite{Yako2006} &  \\
$\sqrt{\left\langle r^{2} \right\rangle_{p} }$ (g.s.) & 4.193 & 5.459 \\
$\sqrt{\left\langle r^{2} \right\rangle_{n} }$ (calc.) & 4.308 & 5.731 \\
$\sqrt{\left\langle r^{2} \right\rangle_{n} }$ (exp.) & $4.26 \pm 0.04$~\cite{Yako2006} & \\
$\delta_{np}$ (calc.) & 0.115 & 0.272 \\
$\delta_{np}$ (exp.) &  0.07 $\pm$ 0.04~\cite{Yako2006} & 0.083 $< \delta_{np} <$ 0.111~\cite{Clark2003} \\
 & & $0.156_{-0.021}^{+0.025}$~\cite{Tamii2011} \\
 & & 0.19 $\pm$ 0.09~\cite{Krasznahorkay1994}  \\
\hline
\end{tabular}
\end{center}
\end{table}
The results obtained in the present study are summarized in Table \ref{tab:nskin}. 
The first and second rows of the table give the values of the sum rule obtained employing the self-consistent 
ground-state mean-field solutions and the calculated SD strengths, respectively. 
The experimental value of the sum rule for $^{90}$Zr \cite{Yako2006} is shown in the third row. 
The difference between the two theoretical results is very small because the model is self-consistent, 
but they both overestimate the experimental value. The radii of proton distributions in the fourth row are 
extracted from the self-consistent ground-state densities, and these values are in excellent agreement 
with data \cite{Vries1987}. In the fifth and sixth rows we include the $rms$ radii of neutron distributions 
in $^{90}$Zr and $^{208}$Pb, calculated from Eq. (\ref{eq:sumrule}) using the proton ground-state radii and 
the calculated SD strength distributions, and the experimental value for $^{90}$Zr, respectively. 
Finally, the calculated and experimental values for the neutron skin thickness  
are given in the last two rows of Table \ref{tab:nskin}.
 
For $^{90}$Zr the calculated sum rule is 9\% larger than the measured value. 
This leads to a neutron rms radius $\sqrt{\left\langle r^{2}\right\rangle_{n}} = 4.308$ fm, 
and neutron skin thickness $\delta_{np} = 0.115$ fm, both at the upper limit of the experimental 
error bars. We note that the relativistic Hartree-Fock + RPA calculation predicts the neutron 
skin thickness $\delta_{np} = 0.092$ fm~\cite{Liang2008}, while Skyrme-based results range 
from $\delta_{np} = 0.055$ fm to $\delta_{np} = 0.106$ fm~\cite{Sagawa2007}. 

There are no experimental values of the total spin-dipole strength in $^{208}$Pb, but several 
measurements of the neutron skin thickness have been reported. A comparison of the measured 
cross section for the isoscalar giant dipole resonance and the DWBA calculation yielded the 
neutron skin thickness $\delta_{np} = 0.19 \pm 0.09$ fm~\cite{Krasznahorkay1994}. 
Microscopic optical potential analyses of intermediate energy elastic proton scattering 
give $\delta_{np} \approx 0.17$ fm~\cite{Karataglidis2002}, 
and $0.083 \text{ fm} < \delta_{np} < 0.111 \text{ fm}$~\cite{Clark2003}. Using the correlation between neutron skin thickness and the isovector dipole polarizability obtained with the Skyrme functional~\cite{Reinhard2010}, the very recent results on polarized proton inelastic scattering at forward angles yield the value  $\delta_{np} = 0.156_{-0.021}^{+0.025} \text{ fm}$~\cite{Tamii2011}. The value obtained from the calculated SD sum rule in the present study is $\delta_{np} = 0.272$ fm, considerably larger than the empirical values. The RHF + RPA model predicts $\delta_{np} = 0.234$ fm~\cite{Liang2008}, and values obtained from various Skyrme-based models range 
between $\delta_{np} = 0.125$ fm and $\delta_{np} = 0.228$ fm~\cite{Sagawa2007}. 
The relatively large neutron skin thickness that we have obtained in the present study is 
peculiar to the relativistic effective interaction NL3 \cite{LKR.97}, characterized by a large 
asymmetry energy. Employing one of the modern relativistic functionals with 
non-linear effective interactions in the isovector channel and lower asymmetry energy as, 
for instance, DD-PC1 \cite{Niksic2008}, the thickness of ground-state neutron skin is calculated: 
$\delta_{np} = 0.088$ fm for $^{90}$Zr, and $\delta_{np} = 0.201$ fm for $^{208}$Pb. 
This type of functionals, however, has not yet been implemented in the RTBA model used in  
this study. 

In summary, the first calculation of charge-exchange spin-dipole
excitations in $^{90}$Zr and $^{208}$Pb, using the particle-vibration
coupling model based on the covariant density functional theory, is
reported. Compared to the RRPA results, the RTBA model 
for particle-vibration coupling leads to pronounced  
fragmentation of the strength distribution for all
three angular-momentum components of the spin-dipole operator. 
A portion of the strength is shifted to higher excitation energy, and the 
corresponding shift of the centroid energy is 2.5 MeV for $^{90}$Zr, 
and 1 MeV for $^{208}$Pb. As a result of particle-vibration coupling a 
high-energy tail of the strength distribution is formed and the strength in this region decreases 
almost linearly with increasing energy, in close agreement with data. 
Furthermore, a model-independent SD sum rule has
been used to determine the neutron $rms$ radii and the thickness of the neutron skin.
The calculated skin thickness $\delta_{np} = 0.115$ fm for $^{90}$Zr and $\delta_{np} = 0.272$ fm 
for $^{208}$Pb, are larger than the available empirical values, and reflect the high asymmetry 
energy of the particular relativistic energy density functional used in the present study.

\bigskip
\leftline{\bf ACKNOWLEDGMENTS}
\noindent 
This work was supported in part by the Helmholtz International Center for FAIR within the framework of the LOEWE program launched by the State of Hesse, the Alliance Program of the Helmholtz Association (HA216/EMMI), 
and by the MZOS - project 1191005-1010.

\end{document}